\documentclass[12pt,epsfig]{article}
\usepackage{graphicx}

\newcommand{\be}[1]{\begin{equation} \label{#1}}
\newcommand{\ee}{\end{equation}}

\newcommand{\bea}[1]{\begin{eqnarray} \label{#1}}
\newcommand{\eea}{\end{eqnarray}}

\newcommand{\refeq}[1]{(\ref{#1})}

\begin{document}

\begin{center}
\Large \bf
 NMR line shapes of a gas of nuclear spin-$\frac{1}{2}$
 molecules in fluctuating nano-containers
\end{center}
\begin{center}

\large \bf
               E.B. Fel'dman \footnote{   E-MAIL: feldman@icp.ac.ru
                                        }
               and M.G. Rudavets
\end{center}
\begin{center}
Institute of Problems of Chemical Physics, \\
         Russian Academy of Sciences, 142432 Chernogolovka,
        Moscow Region, Russia
\end{center}


\begin{abstract}

Reported in this paper is the impact of the fluctuations of the
geometry of the nano-meter gas containers in the medium on
the NMR line shape of the gas inside of the nano-containers.
We calculate exactly the NMR line shape of the gas of
spin-$\frac{1}{2}$ carrying molecules for two typical dynamics of
the nano-container volume and the orientation with respect to the
external magnetic field: (i) for a Gaussian stochastic
dynamics, and (ii) for the regular harmonic vibrations.
For the Gaussian ensemble of static disordered containers having
an infinite correlation time, $\tau_{\sf c} \to \infty $, the
overall line shape is shown to obey a logarithmic low frequency
asymptotics, $ I(\omega) = \mbox{const} \times \ln (\frac{1}{\omega})$, at
$\omega \to 0$, and exponentially decaying asymptotics in a high
frequency domain. For the Gaussian ensemble of the rapidly
fluctuating containers of a finite $\tau_{\sf c}$, the overall line
shape has a bell-shaped profile with $\sim
\omega^{-4}$ far wing behaviour.
In addition, we calculate exactly a satellite structure
of the NMR line shape when the nano-bubbles in a liquid are affected
by the harmonic deformations due to the acoustic waves.

\end{abstract}

\vskip 10mm
PACS nunmers 05.30.-d, 76.20.+q


\newpage
\noindent
\section{Introduction}
\label{sec. 1}

The systems we are studying in this paper are of interest due to the
recent experiments \cite{Baugh} in which NMR responses from the
H$_{2}$ gas confined in the nano-meter porous hydrogenated silicon have
been used for measuring the size and the orientation of the
nano-scale pores \cite{Inag}. The accessibility of these
measurements is relied on the explicit dependence of the line width
of the NMR line shape on the confining volume and on the
orientation of the ordered pores with respect to the strong
external magnetic field. When a gas of spin-carrying molecules is
confined within a nano-meter sized region, the motionally averaged
(effective) dipolar interactions between the nuclear spins take on
the finite values due to the spatial averaging of the dipolar
interactions over the finite ( nano-meter sized ) region, hence,
allowing the estimation of the line width \cite{Baugh} by the
Van-Fleck formula. In addition, the motionally averaged
(effective) dipolar interactions lead immediately to the exactly
solvable effective spin Hamiltonian (see Sec. \ref{sec. 2} below)
allowing the exact calculation of time dependent NMR responses. In
our previous paper \cite{RF} we calculated the time course of the
longitudinal polarization of a gas of nuclear spin-$\frac{1}{2}$
carrying molecules confined in the static nano-containers. The
purpose of the current paper is to extend the developed formalism
for predictions of the NMR line shape in the presence of the
dynamical nano-containers. The example systems we are exploring
here are the cavitation bubbles that could be produced in
experiments like those in sonoluminescence \cite{Bren}, the gas
vesicles emerging in the non-invasive measurements of the liquid flow
by the NMR tomography \cite{Kopt}, \cite{Rio}, the bubbles in blood
\cite{Bru}, the gas bubbles under the nucleation in the course of
liquid-to-gas phase transitions \cite{Klim},
nano-sized free-volume holes of polymers, rubber, etc.
under the elastic deformations, the gas
containers within the vibrating nano-tubes \cite{Dikin},
\cite{Ponch}.

It is common that the NMR signals offer the spectroscopic way of
the measuring of the characteristic sizes and the relaxation times
of the container that surrounds and traps the nuclear spins
\cite{Abragam}, \cite{Kubo}. In studying the NMR signals from the
$N$ spin ensemble with no special symmetry, the exponential with
the number $N$ of allowed unknown basis set of the spin Hamiltonian
is to be incorporated to describe the NMR signals completely.
However, the complexity of the spin Hamiltonian is greatly reduced
when the nuclei are subjected to the fast thermal motion within the
nano-meter container. Under these conditions it arises a vast
difference of the NMR time scale $t_{\sf nmr} = 10^{-4}$ -
$10^{-5}$ sec, that characterizes the flip-flop transitions at the
nearest proton-proton distances, and of the motional time scale,
$t_{\sf m} = 10^{-11}$ - $10^{-12}$ sec, relevant for the round
trip of the hydrogen molecules within the nano-meter
container at room temperature. The presence of a reliable small
parameter
\bea{0.0}
\varepsilon = t_{\sf m}/ t_{\sf nmr} \approx  10^{-7}
\eea
allows to average the underlying dipolar Hamiltonian over the spin
spatial coordinates on the coarse-grained time intervals $\Delta t$
obeying the inequality
\bea{0.1}
t_{\sf m} \ll \Delta t \ll t_{\sf nmr} ,
\eea
giving rise to the motionally averaged spin Hamiltonian $H$ ( see
Eq-s  \refeq{1.1} and \refeq{1.2} of Sec. \ref{sec. 2}) with a
unique spacing independent effective dipolar coupling \cite{Baugh},
\cite{RF}. The exact spectrum of the motionally averaged spin
Hamiltonian $H$ has led to the development of the methods for
characterizing the exact NMR spin dynamics, especially the
non-ergodic spin dynamics and the line shape \cite{RF}.

We mention also the recent studies \cite{Ziel}, \cite{AxelSen} in
the area of the NMR responses from a confined gas in the framework
of the phenomenological Bloch-Torrey equation as well as the NMR
measurements \cite{Gran}, \cite{Kor}, \cite{Web} revealing the
reduction of the atomic mobility with respect to the bulk mobility
due to the confinement effect. Although the previous papers have
been largely concerned with the NMR measurements in the static
nano-containers, the methods developed can be taken over for the
cases of "flexible" walls of nano-containers. The point is that the
dynamics of the surface of a typical bubbling behavior in liquids
occurs at a millisecond time scale \cite{Bren}, i.e. at the same
time scale which is relevant for the NMR spectroscopy. In Section
\ref{sec. 2}, we give a general formalism of the free induction
decay (FID) for an arbitrary moving nano-containers. This is
followed by analysis in Section \ref{sec. 3} of the FID and the
line shape for the Gaussian temporary fluctuations of the volume
and of the orientation of the nano-containers. Our concern in the
Section \ref{sec. 3} is the line shape for a wide range of the
correlation times and the amplitudes of the fluctuations of the
nano-containers. Section \ref{sec. 4} gives the line shape from the
nano-containers subjected to a regular harmonic vibrations at a
single frequency as well as the line shape from the nano-containers
subjected to the harmonic vibrations with the Gaussian
distributions of the frequencies. Finally, Section \ref{sec. 5}
summarizes the major conclusions of the calculations.


\section{ Line shape of a gas within  nano-containers with a time-dependent volume}
 \label{sec. 2}

Consider a gas of $N$ spin-$\frac{1}{2}$ carrying molecules
confined in a moving nano-containers in the strong external
magnetic field $B$. On the coarse-grained time intervals $\Delta t$
\refeq{0.1} the effective spin dynamics is described by the motionally
averaged spin Hamiltonian ,
\bea{1.1}
 H  = \omega_0 I_z +
 \frac{1}{2}D(t) \,
 ( 3 I_{z}^2  - {\bf I}^2 ),
\eea
where the $\omega_0 = \gamma_{\sf p} B$ with $\gamma_{\sf p}$
standing for the proton gyromagnetic ratio, the nuclear spins are
specified by the spin-$\frac{1}{2}$ operators $I_{i\alpha}$,
$i = 1, \ldots, N$, $(\alpha = x, y, z)$, the operators
$ I_{\alpha}= \sum_{i=1}^{N} I_{i\alpha}$,
are referred to as the projections of
the total spin operator onto the $x, y, z$ axes, respectively.
Next, we assigned $ {\bf I}^2 = I_x^2 + I_y^2 + I_z^2 $ to the
square of the total nuclear spin operator. Finally, the motionally
average (effective) dipolar coupling between all ${N \choose 2}$
pairs of the spins in the nano-container is
\bea{1.2}
D(t) = \gamma_{\sf p}^2 \hbar \frac{f(t)}{V(t)} \Bigl( 3\cos^2\theta(t) - 1 \Bigr).
\eea
Here, the $V(t)$ is the volume of the nano-container, the
dimensionless form-factor $f(t)$ depends on the shape of the
nano-container and $\theta(t)$ denotes the time dependent
orientation of the nano-container with respect to the external
magnetic field $\vec B$, see Fig. $1$.
The equivalence of the effective coupling $D$ for all the pairs
of the nuclear spins
is due to the fact that all the nuclei
inside the nano-container are involved in the equivalent fast
thermal motion inside the nano-container
over the long NMR time scale $t_{\sf nmr}$.
In the absence of the nano-container's motion,
the nuclear motion inside the static nano-container gives rise
to the unique motionally averaged spin coupling $D$ \cite{Baugh}, \cite{RF}.
The analytical dependence
of the coupling $D$ on the nano-container's volume $V$ admits
an immediate extension of the coupling $D$ to the case of the time dependent
volume $V$
by invoking the adiabatic framework \refeq{0.1},
i.e. by regarding the motion of the nano-container
to be slow as compared to the fast thermal motion
of the nuclei within the nano-container.

The explicit form of the
function $f(t)$ for the ellipsoidal container is in Ref. \cite{RF}.
For the static containers of the nano-meter volume, $V \sim 10^3 \AA^3$,
the effective coupling in Eq. \refeq{1.2} is evaluated as $10^3$
times smaller than the characteristic flip-flop coupling
($ \sim t_{\sf nmr}^{-1} = \gamma_{\sf p}^2 \hbar/a_0^3 \sim 10^4 - 10^5$
Hz ) of two protons at a nearest separation $ a_0 \sim 1 \AA $.
This reduction of the dipolar coupling is referred to as the fast
motionally narrowing (in $10^3$ times) of the line shape as
compared to the line shape of the static nearest protons.

The line shape is the Fourier transform
of the FID \cite{Abragam},
\bea{1.3}
F(t) =
tr
\left.
\left\{
 \rho(t) I_{-}
\right\}
\right/
tr
\left.\left\{
I_{+}I_{-}
\right\}\right. ,
\eea
where $I_{\pm} = I_{x} \pm \, i I_{y}$ and $\rho(0) = I_{+}$ is the
initial density matrix in the high temperature approximation
\cite{Gold}. The density matrix $\rho(t)$ of the $N$-spin ensemble
in the rotating reference frame obeys the Liouville-von Neumann
(L-vN) equation ($\hbar = 1$)
\bea{1.4}
 i \frac{\partial}{\partial t} \rho =
 \Bigl[
 \frac{1}{2}D(t) \,
 ( 3 I_{z}^2  - {\bf I}^2 ), \rho
 \Bigr]
\eea
In solving L-vN equation \refeq{1.4}, we introduce the phase shift,
\bea{1.5}
\varphi(t) = \frac{1}{2} \int_0^t \, dt' D(t'),
\eea
then, the L-vN  equation \refeq{1.4} is solved to be
\bea{1.6}
\rho(t) =  e^{i \varphi(t) {\mathcal H} } I_{+}  e^{-i \varphi(t) {\mathcal H} }
        =  e^{i 3\varphi(t) (2I_z -1) } I_{+}.
\eea
In deriving \refeq{1.6}, we put
${\mathcal H} =  3 I_{z}^2  - {\bf I}^2 $,
use the commutators
$ \Bigl[  I_{\alpha}, {\bf I}^2 \Bigr] = 0 $
and the identity
\bea{1.7}
e^{i 3\varphi I_z^2  } I_{+}  e^{-i 3\varphi I_z^2 } =
e^{i 3\varphi (2I_z -1) } I_{+}.
\eea
The trace in Eq. \refeq{1.3} is easily performed in the total
occupancy number representation of
$N !/( N_{\uparrow}! N_{\downarrow}!)$-fold
degenerate basis set $\vert N_{\uparrow},
N_{\downarrow} \rangle$, where
$ N = N_{\uparrow} + N_{\downarrow} $
and $ I_z = \frac{1}{2}( N_{\uparrow} - N_{\downarrow} )$.
The trace gives the sought FID for an arbitrary time-dependent
coupling
$D(t)$,
\bea{1.8}
F(t) = \Bigl( \cos \bigl( 3  \varphi (t) \bigr)   \Bigr)^{N-1}.
\eea
On the NMR reasonable time scale $t \le t_{\sf nmr} \sim 10^{-4}$ sec,
we get $ \langle D \rangle t \le  10^{-2}$,
so $ \varphi(t) \ll 1$ and the FID, $F(t)$, of Eq.  \refeq{1.8}
transforms into
\bea{1.9}
F(t) = e^{(N-1)\ln(\cos(3\varphi(t)))   } \simeq
       e^{-\frac{N}{2} (3 \varphi(t))^2 }
\eea
for a large number of spins, $N \gg 1$, in the nano-container. The
FID, $F(t)$, of Eq. \refeq{1.9} involves the effective coupling $D(t)$ of
Eq. \refeq{1.2} as the input parameter to the phase shift $\varphi(t)$
of Eq. \refeq{1.5}. Varying of the function $D(t)$ yields a variety of
the models of the NMR line shape which are commonly discriminated
into the two major groups, viz. the models of the homogeneous or
inhomogeneous line width
\cite{Abragam}, \cite{Kubo}. In the following, we explore the FID
for two dynamical scenarios of the container motion, for a
stochastic Gaussian dynamics (Section \ref{sec. 3}) and for the
regular harmonic oscillations (Section \ref{sec. 4} ).

\section{ Line shape from fluctuating nano-containers}
\label{sec. 3}

When the nano-sized containers are sensitive to the
fluctuations of the environment, we are free to calculate the line
shape by assuming the Gaussian fluctuations of the coupling $D(t)$
\refeq{1.2},
\bea{1.10}
D(t) = \langle D \rangle  +   \delta D (t),
\eea
with the $\delta D (t)$ standing for the
Gaussian random noise characterized by the first two moments
\bea{1.11}
\langle \delta D (t) \rangle = 0, \quad
\langle \delta D (t_1) \delta D (t_2)  \rangle =
\langle ( \delta D )^{2}  \rangle  \,  C( | t_1 - t_2 |),
\eea
where
$\langle ( \delta D )^{2} \rangle$
is the variance of the
fluctuations and the $C(t)$ denotes the correlation function, for
example,
$
C(t) = \exp(-t/\tau_{\sf c}) $, with $\tau_{\sf c}$ being the
correlation time. The averaging of the function $F(t)$ over the
Gaussian fluctuations $\delta D (t)$ is carried in two steps:
first, we rewrite the FID, $F(t), of Eq. $ \refeq{1.9} by introducing the
Gaussian parameterization,
\bea{1.12}
F(t) = \int_{-\infty}^{+\infty}
\frac{\,dx}{\,\sqrt{\pi}}
\,
e^{ -x^2 - 3 i x \sqrt{2N} \varphi(t)},
\eea
this is followed by the second step of
applying the formula
for the averaging, see e.g. \cite{Abragam}, \cite{Kubo},
of the function $F(t)$ of Eq. \refeq{1.12}
over the random Gaussian process $\delta D(t)$
entering the phase $\varphi(t)$  of Eq. \refeq{1.5},
\bea{1.13}
\left \langle
\exp \left( -i \kappa \int_0^t \delta D(t') dt' \right)
\right\rangle_{\delta D}   &  =  &
\exp \Bigl(
-  \kappa^2
\langle ( \delta D )^{2}  \rangle  T^2(t)
     \Bigr),                                 \nonumber  \\
T^2(t) & = &   \int_0^t (t- t') \,  C(t')\, dt'  ,
\eea
with the constant $\kappa = 3 x \sqrt{N/2}$.
Averaging by Eq-s. \refeq{1.13}, \refeq{1.12} gives the sought FID
\bea{1.14}
F(t) = \frac{
             \exp\Bigl(-\frac{ t^2 \nu^2/4       }
                       { 1 + \alpha \nu^2 T^2(t) }
                 \Bigr)
            }
            { \sqrt{1 + \alpha \nu^2 T^2(t)}
            },
           \quad
\alpha =
\frac{\langle (\delta D)^2 \rangle}{\langle D \rangle^2 },
\quad
\nu = 3 \langle D \rangle  \sqrt{ \frac{N}{2} } .
\eea
As it stands, the FID, $F(t)$, of Eq. \refeq{1.14} encodes an information on
the mean volume and the mean orientation of the cavity with respect
to the external magnetic field as well as on the fluctuations of
the cavity.

If the fluctuations were absent ($\alpha = 0$ in Eq.  \refeq{1.14})
then the FID, $F(t)$, of Eq. \refeq{1.14} forms the line shape,
\bea{1.15}
I(\omega) = \frac{1}{\pi}\int_{0}^{\infty}
            F(t) \cos{\omega t} \,dt
\eea
of the Gaussian type,
$
I(\omega) =
\frac{1}{ \nu \sqrt{\pi} }
\exp
\Bigl(
-
\frac{  \omega^2 }
     {\nu^2}
\Bigr)
$.
Fluctuations of the cavity ($\alpha \neq 0$ in Eq. \refeq{1.14}),
result in the broadening of the line
shape so that the second moment, see e.g. \cite{Abragam}, reads
\bea{1.16}
M_2 = - \left. \frac{d^2 F(t)}{d(t)^2}  \right|_{t=0}
    =   \frac{9N}{4}\left[
\langle D \rangle^2 + \langle (\delta D)^2 \rangle
                  \right].
\eea
In deriving the second moment, use is made of the expression $T^2$
in Eq. \refeq{1.13} and the property $C(0) = 1$.

To make the calculations of the line shape $I(\omega)$ more explicit, we take
the exponential correlation function,
$
C(t) = \exp(-t/\tau_{\sf c})
$. It follows
\bea{1.16.2}
T^2(t) = \tau_{\sf c}^2 \,
                     \Bigl(
         \exp(-t/\tau_{\sf c}) + \frac{t}{\tau_{\sf c}} -1
                     \Bigr).
\eea
In the analysis of the FID, $F(t)$, of Eq.  \refeq{1.14} with the function
$T^2(t)$ of Eq. \refeq{1.16.2}, it seems very useful to consider the
temporary fluctuations on two very different time scales.

When
$
\tau_{\sf c}^2 \langle (\delta D)^2 \rangle \ll 1
$,
the function $T^2(t)$ is
$T^2(t) = \tau_{\sf c} t $ for times $t \gg \tau_{\sf c}$,
so that the FID of Eq. \refeq{1.14} reads
$
F(t) = \exp\Bigl( -\frac{1                    }
                        { 4\alpha\tau_{\sf c} } t
          \Bigr)/
            { \sqrt{1 + \alpha \tau_{\sf c} \nu^2 t } },
$
and its Fourier transform of Eq. \refeq{1.15} admits the following representation
\bea{1.16.3}
I(\omega) = \frac{1}{\alpha \sqrt{\pi} \tau_{\sf c} \nu^2}
            \mathop{\rm Re}
            \Bigl(
            \frac{e^{z}    }
                 {\sqrt{z} }
            \mathop{\rm erfc}(\sqrt{z})
            \Bigr),
            \quad z = (2\alpha \tau_{\sf c} \nu)^{-2} +
            i \omega (\alpha \tau_{\sf c} \nu^2)^{-1} ,
\eea
in terms of the function $\mathop{\rm erfc}$ \cite{AS}. The
function $I(\omega)$ of Eq. \refeq{1.16.3} has a bell-shaped profile with
an intermediate Lorentzian asymptotics
$I(\omega) = \frac{1}{\pi}\Gamma /(\Gamma^2 + \omega^2)$
where
$\Gamma = (4\alpha \tau_{\sf c})^{-1}$
( see in \cite{AS} the asymptotics of the function
$\mathop{\rm erfc}(\sqrt{z})$ at $\vert \sqrt{z} \vert \gg 1$ or
$\tau_{\sf c} \to 0$ ). Far wing calculations of the line shape
$I(\omega)$ require the FID
$F(t)$ at $0 \leftarrow t < \tau_{\sf c}$,
which is provided by the function $T^2(t)$ of Eq. \refeq{1.16.2} at
$t \to 0$. The sought asymptotics of the line shape $I(\omega)$
can be evaluated as
\bea{1.16.3.2}
I(\omega) \sim  \frac{9N}{4 \pi }
          \omega^{-4}
          \langle (\delta D)^2 \rangle \frac{1}{\tau_{\sf c}},
          \quad \omega \to \infty
\eea
on integrating the line shape $I(\omega)$
of Eq. \refeq{1.15} four times by parts and employing the derivatives
$F'(0) = 0$,
$
C'(0) = - \tau_{\sf c}^{-1}
$.

On the other hand, when
$
\tau_{\sf c}^2 \langle (\delta D)^2 \rangle \gg 1
$,
the function $T^2(t)$ becomes
$T^2(t) = t^2/2$ for
$0 \leq t \leq \tau_{\sf c} \to \infty$.
The function $T^2(t)$  leads immediately to the slowing down of the FID
of Eq. \refeq{1.14} in the form
$
F(t) = \exp\Bigl( -\frac{1        }
                        { 2\alpha }
          \Bigr)/
            { \sqrt{1 + \frac{1}{2} \alpha \nu^2 t^2 } }
$. Consequently, one is left with the line shape
\bea{1.16.4}
I(\omega) & = &\frac{ e^{-\frac{1}{2\alpha}} }{\pi}
               \int_{0}^{\infty}
               \frac{ \cos{\omega t} \,dt                        }
                    {\sqrt{1 + \frac{1}{2} \alpha \nu^2 t^2 } } =
          \frac{ e^{-\frac{1}{2\alpha}} }{\pi \nu}
          \sqrt{ \frac{2}{\alpha} } \,\,
           K_0
             \Bigl(
          \frac{ \omega \sqrt{2}}{ \nu \sqrt{\alpha} }
             \Bigr) =
                                              \nonumber          \\
          & = &
          \frac{ e^{-\frac{1}{2\alpha}} }{\pi \nu}
          \sqrt{ \frac{2}{\alpha} } \,\,
             \Bigl[
          \ln \Bigl( \frac{ \nu \sqrt{\alpha} }{ \omega \sqrt{2} } \Bigr) + O(1)
             \Bigr], \quad \omega \to 0,
\eea
and
\bea{1.16.5}
I(\omega) =
\frac{ e^{-\frac{1}{2\alpha}}              }
     {\sqrt{ \pi \nu \sqrt{ 2\alpha }  }   }
\frac{  1             }
     { \sqrt{\omega}  } \,
     \exp                     \Bigl(
                                -\frac{\omega \sqrt{2} }{ \nu \sqrt{\alpha} }
                              \Bigr)
                          , \quad \omega \to \infty.
\eea
In deriving the $I(\omega)$ of Eq.-s \refeq{1.16.4} and \refeq{1.16.5}, use is
made of the integral representation of the modified Bessel function
$K_0(x)$ and its asymptotics \cite{AS}. In order to shed light on
the slowing down of the FID $F(t)$ at $t\to \infty$ (and, thus, on
the logarithmic singularity of the line shape $I(\omega)$ at
$\omega \to 0$ in eq. \refeq{1.16.4}), we regard a disordered
almost static (frozen at $\tau_{\sf c} \to \infty$) distribution of
the nano-containers with the  various volumes. The FID from an
individual nano-container is described by the Gaussian-in-time
function with the relaxation rate $\nu = 3 D \sqrt{ N/2 }$. The
main contribution to the sum of the individual FID's weighted by
the Gaussian probability density
$
(2\pi\langle D^2 \rangle)^{-1/2} e^{ -D^2 / 2\langle D^2 \rangle }
$
of the static fluctuations of the coupling $D$ comes from the
fluctuations having the $ D = 0$, i.e. from the containers with a
large volume or oriented at the magic angle
$\mathop{\rm arccos}(1/\sqrt{3} )$ ( see the expression
for the coupling $D$ of Eq. \refeq{1.2}).
The FIDs from the large cavities are slowly damping providing a
slow damping of the overall signal
$
F(t) \sim 1/t
$
at $t \to \infty$ rather than the Gaussian-in-time
asymptotics. Notice that the slowing down of the overall FID from
the Gaussian ensemble of the static fluctuating nano-containers
akin the slowing down of the unimolecular decay on the static
disordered traps \cite{BV}, \cite{GP}. Fig. $2$ is aimed to show
the appearance of a low frequency singularity of the line shape
while increasing $\tau_{\sf c}$ to infinity. The line shape for
fixed $\tau_{\sf c}\nu = 10$ at the various $\alpha$ is displayed on
Fig. $3$. In addition to having the broad shape, the line shape
$I(\omega)$ shows the singularity at the zero frequency asymptotics
at the large fluctuations $\alpha$.

It is worth to gather a small
number of characteristic quantities involved in the Figure $3$; for
$N = 500$ spin-$\frac{1}{2}$ molecules
( $\gamma_{\sf p}^2 \hbar = 2\pi \cdot 120$ Hz $\cdot$ nm$^3$)
within $V = 45 $ nm$^3$
nano-container and the form-factor $f(t) \sim 2$ of Eq. \refeq{1.2},
the motionally average dipolar interaction is
evaluated as $\langle D \rangle = 2\pi \cdot 5.3$ Hz, thus,
$\nu = 0.25 \cdot 10^3$ Hz
and $\tau_{\sf c} = \nu^{-1} \simeq 4 \cdot 10^{-3}$ s.

The  lessons drawn from this Section are that the fluctuations of
the nano-containers give rise to the deviation of the line shape
from the standard Gaussian and the Lorentzian shapes,
and that for almost static disordered nano-containers
at
$
\tau_{\sf c}^2 \langle (\delta D)^2 \rangle \gg 1
$,
the line shape $I(\omega)$ gets narrower at $\omega \to 0$ and
broader at $\omega \to \infty$ as compared to the bell-shaped profile occurring for
frequently fluctuating nano-containers
at
$
\tau_{\sf c}^2 \langle (\delta D)^2 \rangle \ll 1
$
.

\section{ Line shape from vibrating nano-containers}
\label{sec. 4}

Acoustic waves in the liquid surrounding the nano-bubbles can
induce a synchronized harmonic vibrations of the nano-bubble
volumes and their orientations \cite{Bren}, thus, affect the NMR line
shape if the acoustic waves are at the NMR relevant frequency
domain $1 - 10$ kHz. The same physical picture of the NMR responses
should appear for a gas within the vibrating nano-tubes
\cite{Dikin}, \cite{Ponch}. For all these vibrating
nano-containers, we can regard the coupling $D(t)$ of Eq. \refeq{1.2} to
be a harmonic function of the time,
\bea{2.8}
D(t) = \langle D \rangle (1 + \varepsilon \cos(\Omega t))
\eea
with parameter
$\varepsilon < 1$ assuming a weak vibrations of the nano-bubble
volumes and the orientations.
On the NMR time scale $t \sim 10^{-3}$ sec,
the  phase shift $\varphi(t)$ of Eq.  \refeq{1.5} becomes
\bea{2.9}
\varphi(t) = \frac{1}{2}\langle D \rangle
(t + \frac{\varepsilon}{\Omega} \sin(\Omega t)),
\eea
so that
$ \varphi(t) \ll 1$
and the signal $F(t)$ of Eq. \refeq{1.8}
again transforms into the
$
F(t) = e^{-\frac{1}{2}  (3 N \varphi(t))^2 },
$
( see Eq. \refeq{1.9}).
For a weak vibrations, it is enough to expand
the function $F(t)$ in the powers of the parameter $\varepsilon$
keeping only the terms up to $\varepsilon^2$ and linear in a small
factor $\nu^2 = \frac{9}{2} \langle D \rangle^2 N $,
( by  Eq. \refeq{1.2}, the coupling
$\langle D \rangle \sim 1/N $), giving
\bea{2.11}
F(t) = e^{-t^2 \nu^2/4}
\Bigl(
1 -  \varepsilon   \frac{ \nu^2 t}{2\Omega} \sin(\Omega t)
  -   \varepsilon^2  \Bigl( \frac{\nu}{2\Omega} \Bigr)^2    \sin^2(\Omega t)
\Bigr) .
\eea
The Fourier transformation brings the signal $F(t)$ of Eq. \refeq{2.11}
into the line shape
\bea{2.12}
I(\omega, \Omega) & = & \frac{ 1 }{ \nu \sqrt{\pi} }
   e^{-\frac{\omega^2}{\nu^2}}
 + \sum\limits_{i = 1, 2} \Bigl(
                I_{i}(\omega, \Omega) +  I_{i}(-\omega, \Omega)
                          \Bigr),
                                      \nonumber          \\
I_{1}(\omega, \Omega) & = & -\frac{  \varepsilon       }
                                  { 2\nu\sqrt{\pi}  }
\Bigl(
1 + \frac{ \omega }{\Omega}
\Bigr)
e^{-\frac{(\omega + \Omega)^2 }{\nu^2}},
                                      \nonumber          \\
I_{2}(\omega, \Omega) & = &
\frac{\varepsilon^2  }{4 \nu \sqrt{\pi} }
\Bigl( \frac{ \nu }{2 \Omega} \Bigr)^2
e^{-\frac{(\omega + 2\Omega)^2 }{\nu^2}}.
\eea
In the absence of the bubble vibrations $(\varepsilon = 0)$, the
line shape $I(\omega, \Omega)$ exhibits the motionally narrowed
peak at frequency $\omega=0$ with the line width $2\nu$. Weak
bubble vibrations at the single frequency $\Omega$ leads to the
appearance of the symmetric satellite pairs of the line shape
$I(\omega, \Omega)$ \refeq{2.12} at the multiple frequencies
$\omega = \pm \Omega,\, \pm 2\Omega$. By accounting for the $n$-th
term in the powers of the amplitude $\varepsilon$, the satellite
pairs at frequencies $\omega = \pm n \Omega$, $ n = 3, 4, \dots$
arise.

Now, let an ensemble of many individual gas bubbles is spread over
the liquid having a random local vibrational frequency $\Omega$ due
to an intimate fluctuations of the liquid. Under these conditions,
the nuclei belonging to  different bubbles are (indirectly)
subjected to a different local vibrational frequency resulting to
the inhomogeneous broadening of the NMR spectrum. We think of the
ensemble of the bubbles as a single bubble affected by the
vibrations with a continuous Gaussian distribution of frequencies
\bea{2.13}
D(\Omega) = A_0 \Omega^2 e^{-\frac{(\Omega - \Omega_0)^2 }{\Delta^2}}
\quad
\mbox{with}
\quad
A_0 = \frac{1}{
\sqrt{\pi} \Delta
\Bigl(
\frac{1}{2} \Delta^2 + \Omega_0^2
\Bigr)
             }
\eea
ensuring the normalization
$\int_{-\infty}^{\infty} \, d \Omega D(\Omega) = 1$.
The pre-factor $\Omega^2$ in Eq. \refeq{2.13}
is taken for ease of performing the averaging of the line shape
$I(\omega, \Omega)$ of Eq. \refeq{2.12} over the distribution $D(\Omega)$ of Eq. \refeq{2.13},
\bea{2.14}
\langle I(\omega) \rangle  =
\int_{-\infty}^{\infty}
\, d \Omega  D(\Omega) I(\omega, \Omega),
\eea
yielding
\bea{2.15}
\langle I(\omega) \rangle  & = &
  \frac{1 }{ \nu \sqrt{\pi}  }
   e^{-\frac{\omega^2}{\nu^2}}
 + \sum\limits_{i = 1, 2} \Bigl( G_{i}(\omega) + G_{i}(-\omega) \Bigr),
                                      \nonumber          \\
                    G_{1}(\omega) & = &
 -\frac{\varepsilon }{2}
  \frac{ A_0 \Delta^3             }
       {\nu (1 + \delta^2 )^{3/2} }
\Bigl[
\frac{1}{2} +
\frac{          ( \Omega_0 - \omega \delta^2 ) }
     { \Delta^2 ( 1  +  \delta^2     )         }
    \Bigl(
    \Omega_0 + \omega
    \Bigr)
\Bigr]
e^{ -\frac{(\omega + \Omega_0)^2 }{\nu^2 + \Delta^2} },
                                      \nonumber          \\
                   G_{2}(\omega) & = &
\frac{\varepsilon^2}{16}
\frac{ A_0 \nu \Delta       }
     {(1 + \delta^2 )^{1/2} }
e^{ -\frac{(\omega + 2\Omega_0)^2 }{\nu^2 + 4\Delta^2} },
\eea
with $\delta = \Delta/\nu$. The overall line shape $\langle
I(\omega) \rangle$ of Eq. \refeq{2.15} is still narrow at $\omega =0$ and
reveals the two symmetric satellite pairs at frequencies
$\omega = \pm \Omega_0,\, \pm 2\Omega_0$ having a broad line width
$2\sqrt{\nu^2 + \Delta^2}$
and $2\sqrt{\nu^2 + 4\Delta^2}$, respectively, see
Fig. $4$.

In general, the satellite pair of the amplitude
$\varepsilon^n$ is described (to within the pre-exponential factor)
by the Gaussian shape with the line width
$2 \sqrt{\nu^2 + n^2 \Delta^2} \sim 2 n\Delta $,
i.e. $2n$ times larger than the
dispersion, $ \Delta $, of the frequencies in the spectral density
$D(\Omega)$ of Eq. \refeq{2.13}. Thus, the position of the $n$-th
satellite pair at $\omega = \pm n \Omega_0$ and the broadening of
the $n$-th satellite, $2n\Delta $, provide the NMR spectroscopic
characterization of the nano-bubble vibrations happening at the
mean vibrational frequency $\Omega_0$ and with the dispersion of
the frequencies $\Delta $.


\section{Conclusion}
\label{sec. 5}

The focus in the paper is on the exact NMR line shape theory of a
gas of spin-$1/2$ carrying molecules confined within the
fluctuating nano-containers. Two typical dynamics of the
nano-containers was treated, viz. the Gaussian stochastic dynamics
and the regular harmonic vibrations.

$(1)$. Of the variety of the Gaussian random fluctuations of the
nano-containers, the most striking fluctuation effect on the NMR
line shape is due to the fluctuations at the large correlation times,
$\tau_{\sf c} \to \infty$, (for almost frozen disordered ensemble
of the various nano-containers) and at the large amplitudes of the
fluctuations of the volume and orientation of the nano-containers.
Under these conditions, the NMR line shape behaves as
$I(\omega) = \mbox{const} \times \ln\frac{1}{\omega}$
at $\omega \to 0$ and
exponentially decaying at the large frequencies, $\omega \to \infty$.
Alternatively, when the conditions are specified by the small
correlation times, $\tau_{\sf c} \to 0$, or at the small
amplitudes, $\alpha$,
of the Gaussian fluctuations of the nano-containers, then the line
shape has the bell-shaped profile with the power law $\sim
\omega^{-4}$ at far wings. The line width and its precise shape
specify the mean volume, the mean orientation of the cavities as
well as the deviation of the volumes and the orientations from the
mean values.

$(2)$. If the driving sources of the vibrations support the
harmonic vibrations of the bubble volumes and of the orientations at a
single frequency $\Omega$, then the line shape has the spike
satellite pairs with a narrow line width $2\nu$ at the frequencies
$\omega = \pm n \Omega$, $n = 1, 2,\dots$ around the central spike
at $\omega = 0$. For the Gaussian distribution of driving
frequencies with the mean $\Omega_0$ and the dispersion $ \Delta$,
the central spike at $\omega =0$ remains to be narrowed with the line
width $2 \nu$, however, the satellite pairs at $\omega = \pm n
\Omega_0$, are subjected to broadening in the way that the $n$-th
pair has the line width $\sim 2 n\Delta $.

The upshot is that the paper demonstrates how the fluctuation
dynamics in the medium can be characterized by the NMR spectroscopy
of the gas within the fluctuating nano-containers.


\vskip 5mm

\noindent

{\bf Acknowledgments}

\vskip 5mm

Thanks are expressed to I.I. Maximov for the help in preparing the
manuscript.
Financial support was provided
by the Russian Foundation of Basic Research (RFBR No. 04-03-32528).




\newpage
\noindent
{\large  \bf
Captions to figures.
}
\vskip 5mm

Fig. $1$.
Cartoon of the two positions of the nano-container which is moving
in a liquid undergoing the deformations of the volume and/or the variation of the
orientation $\theta(t)$ with the time.
The nano-container confines the gas of nuclear spin-$\frac{1}{2}$ molecules
uniformly spread inside the nano-container.

Fig. $2$.
The line shapes $I(\omega)$ of Eq. \refeq{1.15} are calculated by
the Fourier transform of the FID $F(t)$ of Eq. \refeq{1.14} with the
function $T^2(t)$ of Eq. \refeq{1.16.2} for the various values of
the combination $\tau_{\sf c}\nu$.
The parameters $\nu$ and the $\alpha = 1$ are from  Eq. \refeq{1.14}.

Fig. $3$.
The line shapes $I(\omega)$ of Eq. \refeq{1.15} for the FID $F(t)$ of Eq. \refeq{1.14}
with the function $T^2(t)$ of Eq. \refeq{1.16.2}
at the fixed $\tau_{\sf c}\nu = 10$,
but the amplitude of the fluctuations are allowed to vary from $\alpha = 0.01$
to $\alpha = 100$.

Fig. $4$.
The absolute value of the homogeneous (dashed-dot) and inhomogeneous (solid) NMR
line shapes of
the vibrating bubbles with the  nuclear spin-$\frac{1}{2}$
molecules inside. The absolute value of the
inhomogeneous line shape $|\langle I(\omega)\rangle |$
from Eq. \refeq{2.15} is shown  for the
parameters $\varepsilon = 5$, $\Delta = 2$, $\Omega_0 = 2\pi$, with
all the frequencies being in the units of the frequency $\nu$ of Eq. \refeq{1.14}.


\begin{thebibliography}{99}
\bibitem{Baugh} J. Baugh, A. Kleinhammes, D.Han, Q. Wang, and Y. Wu,
                   Science  294 (2001) 1505.
\bibitem{Inag}  S.Inagaki, S.Guan, T. Ohsuna and O. Terasaki,
                  Nature 416 (2002) 304.
\bibitem{RF} E.B. Fel'dman and M.G. Rudavets ,
                  JETP  98 (2004) 207;
                  E.B.Fel'dman and M.G.Rudavets,
                      ArXive e-print quant-ph/0306055.
\bibitem{Bren} M. P. Brenner, S. Hilgenfeldt and D.Lohse,
                     Rev. Mod. Phys.,  74 (2002) 425.
\bibitem{Kopt} I.V. Koptyug and R.Z. Sagdeev,
                    Russ. Chem. Rev.  71 (2002) 789.
\bibitem{Rio} F.Rioual, T. Biben and C.Misbah, ArXive e-print
                                         physics/0401159.
\bibitem{Bru} E.A. Brujan,
                   Europhys. Lett.,  50 (2000) 437.
\bibitem{Klim} V. V. Klimov and V. S. Letokhov,
                     Chem. Phys. Lett. 301 (1999) 441.
\bibitem{Dikin} D.A. Dikin, X. Chen, W. Ding, G.J. Wagner and R.S. Ruoff,
                            J. Appl. Phys.  93 (2003) 226.
\bibitem{Ponch} P. Poncharal, Z.L. Wang, D. Ugarte
                 and W.A. de Heer, Science 283 (1999) 1513 .
\bibitem{Abragam} A. Abragam, The Principles of Nuclear Magnetism,
                     Clarendon Press, Oxford,  1961.
\bibitem{Kubo} R. Kubo,
                  Adv. Chem. Phys.  15 (1969) 101.
\bibitem{Ziel} L. J. Zielinski, P.N. Sen,
               J. Chem. Phys.  119 (2003) 1096.
\bibitem{AxelSen} S. Axelrod and P. N. Sen,
                  J. Chem. Phys. 114 (2003) 6879.
\bibitem{Gran} S. Granic,                  Science  253 (1991) 1374.
\bibitem{Kor}  J.-P. Korb, L. Malier, F. Cros, S. Xu, J. Jonas,
                     Phys. Rev. Lett. 77 (1996) 2312.
\bibitem{Web} M. Weber, A. Klemm, R. Kimmich,
                 Phys. Rev. Lett.  86 (2001) 4302.
\bibitem{Gold} M. Goldman, Spin Temperature and Nuclear Magnetic
                           Resonance in Solids,
                           Clarendon Press, Oxford,  1961.
\bibitem{AS}    M. Abramowitz and A.I. Stegun,
                 Handbook of Mathematical Functions,
                 Dover, New York, 1965.
\bibitem{BV} B. Ya. Balagurov and V. G. Vaks, Zh. Eksp. Teor. Fiz.
                  65 (1973) 1939.
\bibitem{GP} P. Grassberger and I. Procaccia, J. Chem. Phys.
                  77 (1982) 6281.
\end{thebibliography}
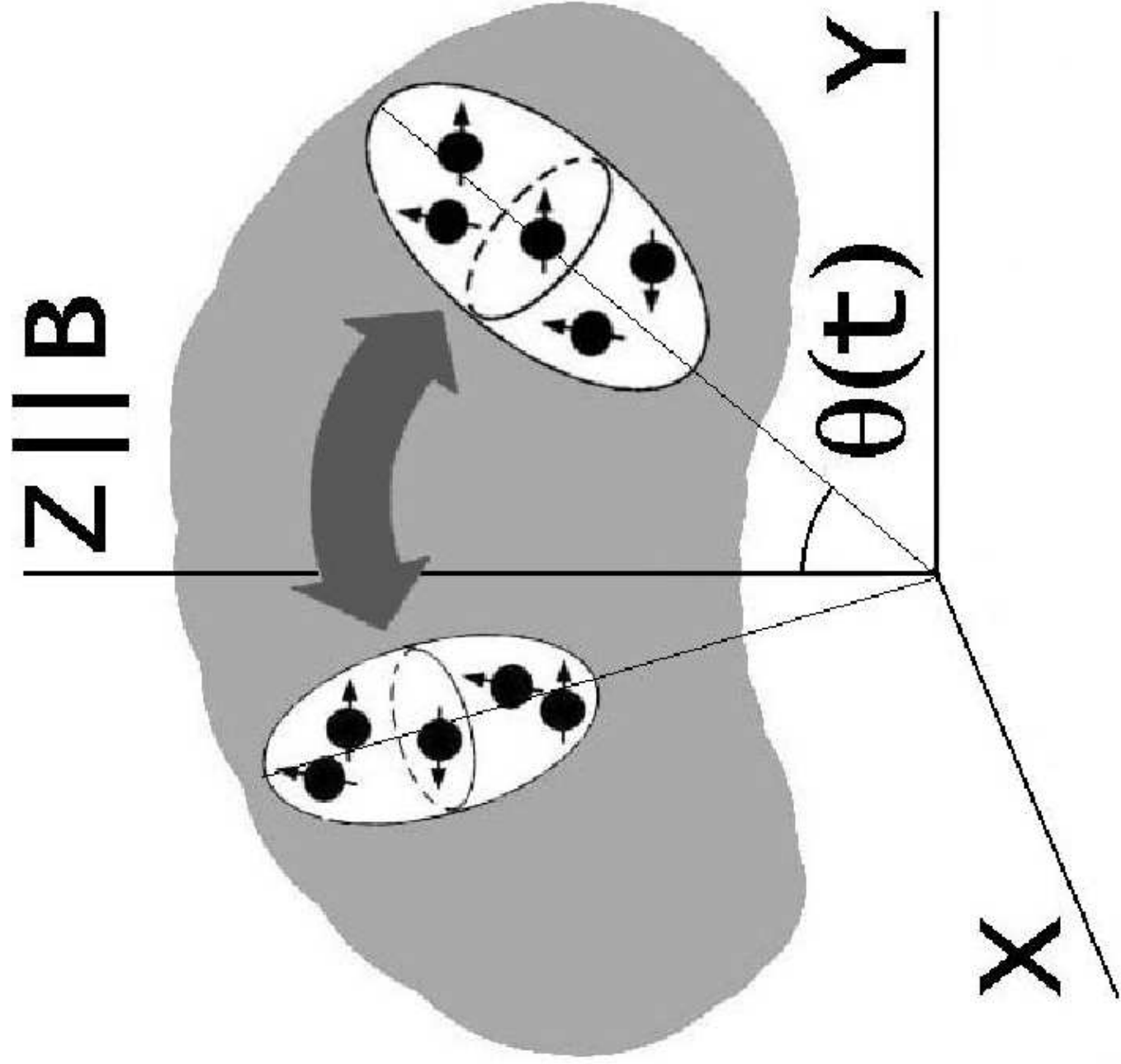
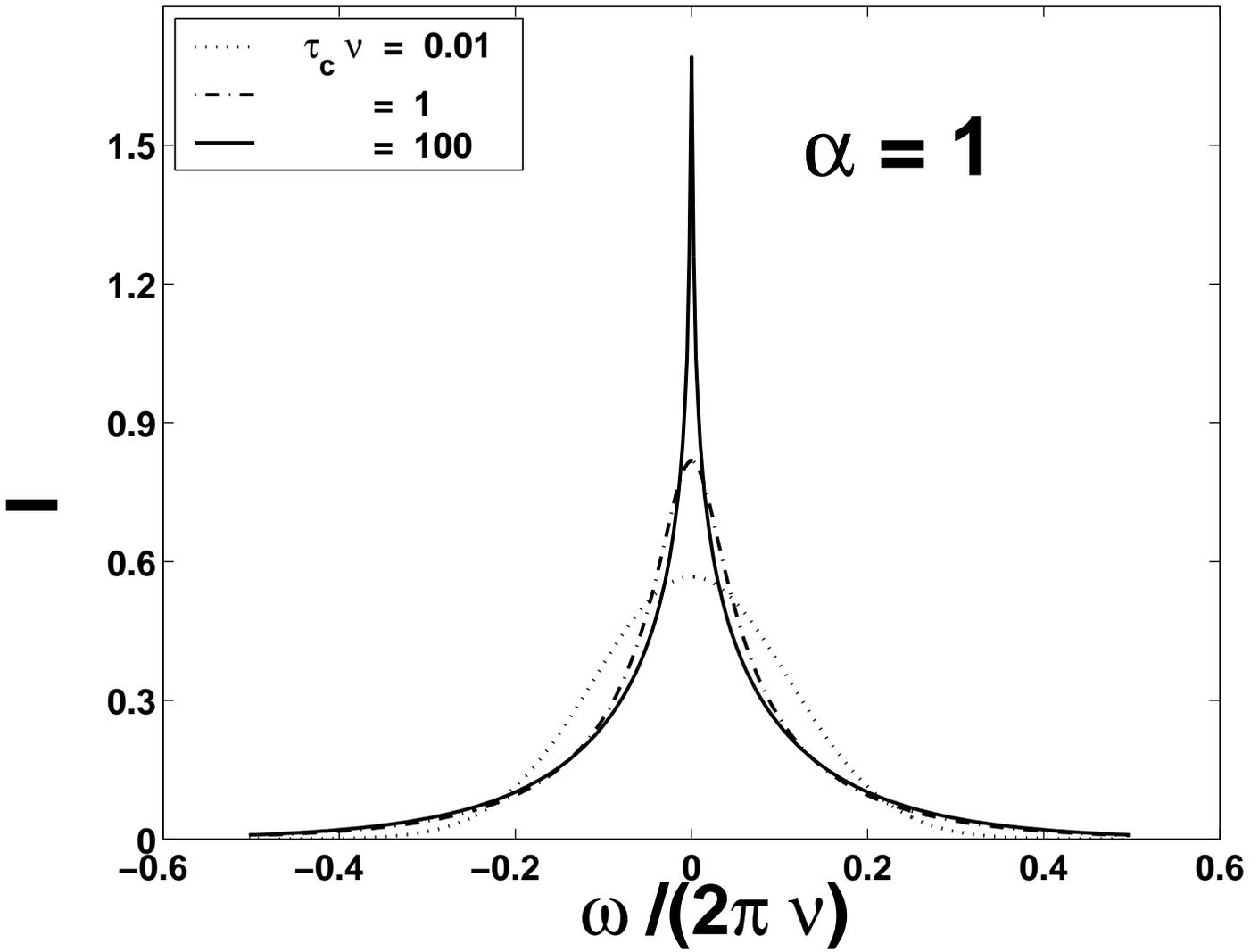
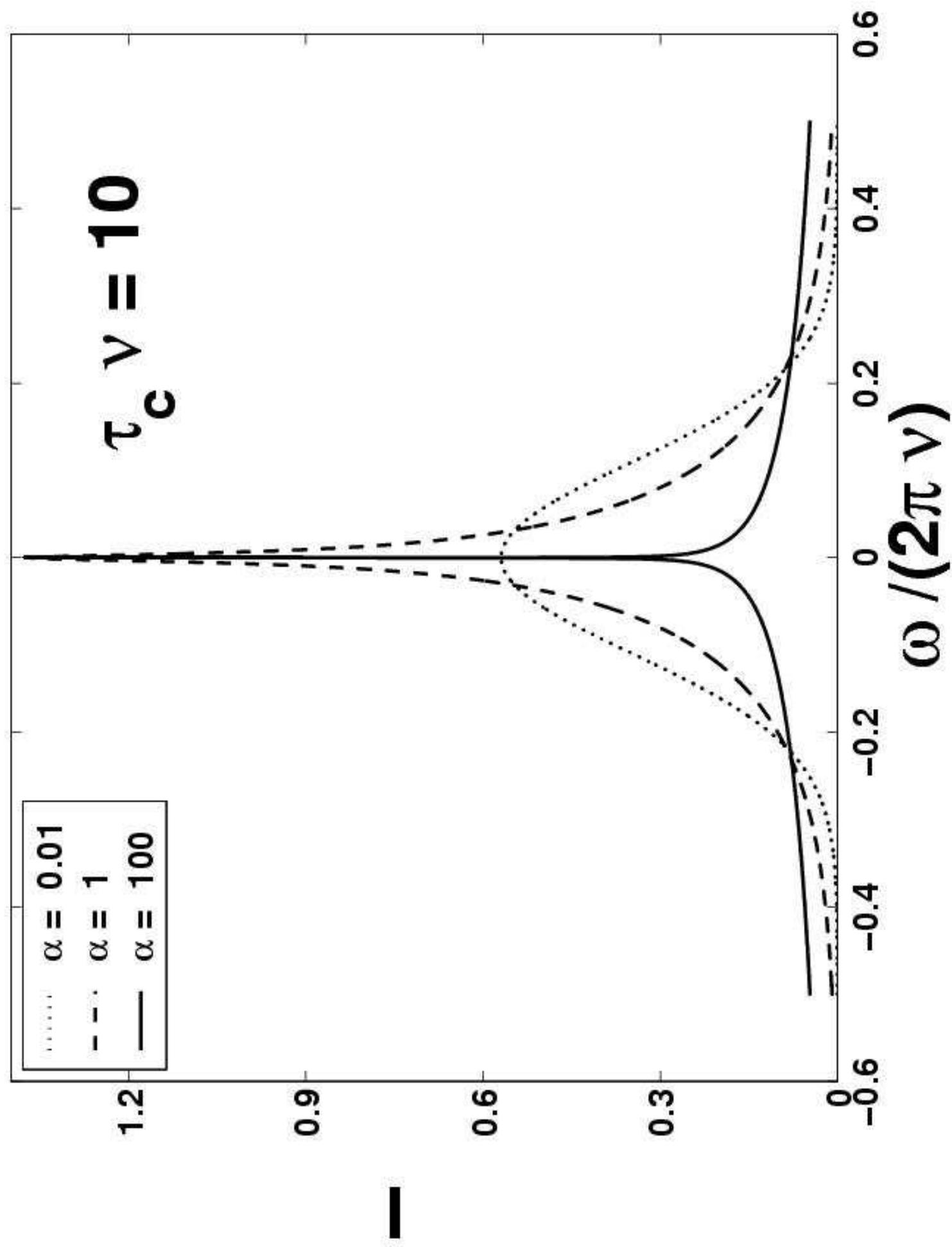
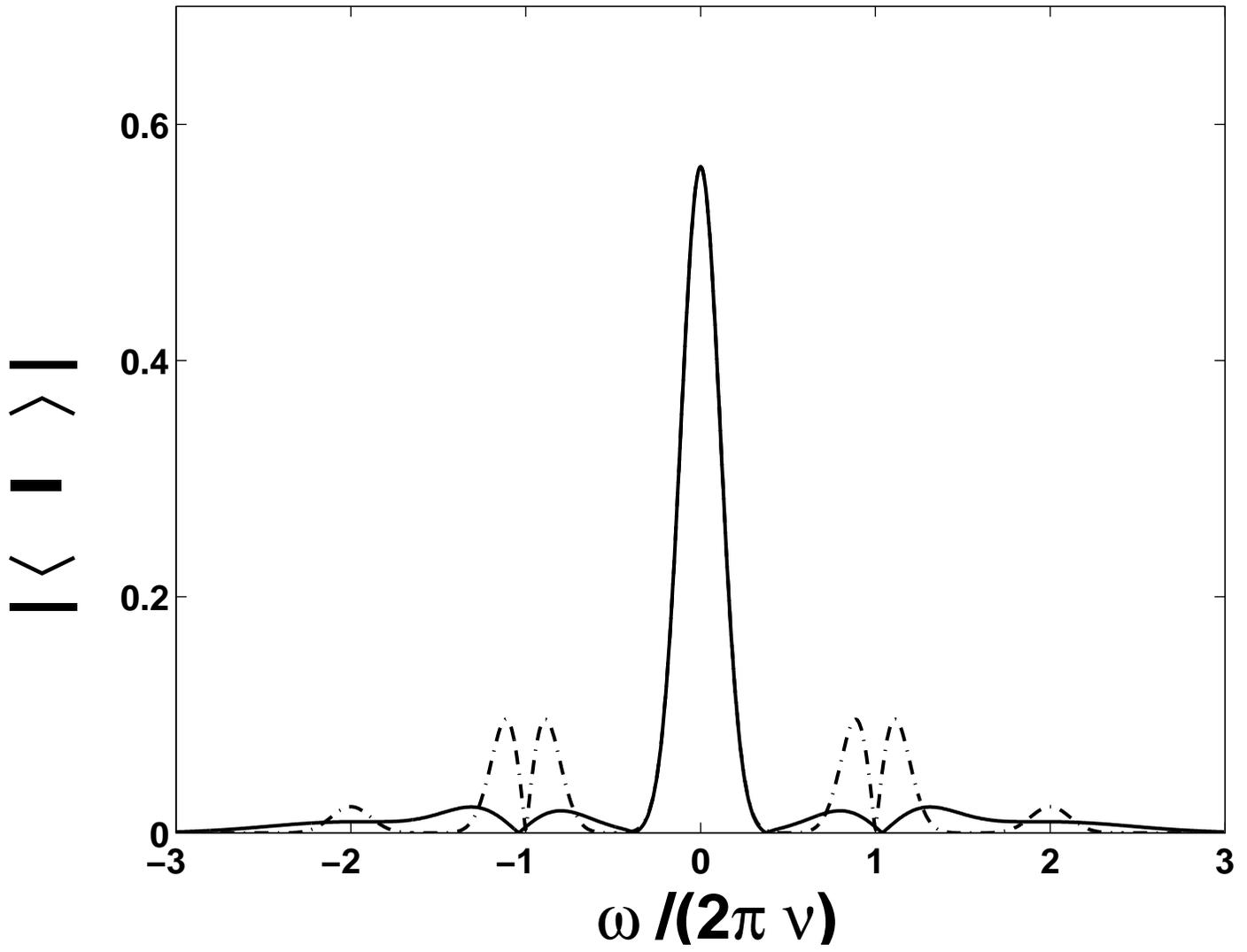
\end{document}